\documentclass[showkeys,onecolumn]{revtex4}
\usepackage{amsfonts}
\usepackage{amssymb}
\usepackage{amsmath}
\usepackage{graphicx}
\usepackage{epsfig}

\begin{document}

\title{Impurity-induced gap system as a quantum data bus for quantum state
transfer}
\author{Bing~Chen$^{1}$}
\email{chenbingphys@gmail.com}
\author{Yong~Li$^{2}$}
\author{Z.~Song$^{3}$}
\author{C.-P.~Sun$^{2}$}
\affiliation{$^{1}$ Department of Applied Physics, College of Electronics, Communication
and Physics, Shandong University of Science and Technology, Qingdao 266510,
China}
\affiliation{$^{2}$ Beijing Computational Science Research Center, Beijing 100084, China}
\affiliation{$^{3}$ School of Physics, Nankai University, Tianjin 300071, China}
\date{\today }

\begin{abstract}
We introduce a tight-binding chain with a single impurity to act as a
quantum data bus for perfect quantum state transfer. Our proposal is based
on the weak coupling limit of the two outermost quantum dots to the data
bus. First show that the data bus has an energy gap between the ground
and first-excited states in the single-particle case induced by the impurity
in the single particle case. By connecting two quantum dots to two sites of
the data bus, the system can accomplish a high-fidelity and long-distance
quantum state transfer. Numerical simulations were performed for a finite
system; the results show that the numerical and analytical results of the
effective coupling strength agree well with each other. Moreover, we study
the robustness of this quantum communication protocol in the presence of
disorder in the couplings between the nearest-neighbor quantum dots. We find
that the gap of the system plays an important role in robust quantum state
transfer.
\end{abstract}

\maketitle

\section{INTRODUCTION}

The transfer of quantum states from one quantum unit of a quantum computer
to another is of fundamental importance in quantum information science.
Recently, in view of the great potential of a physical realization of
the quantum computer, attention is being paid to the problem of the transfer
of quantum information in a solid-state system.

Recently, spin systems have been proposed as a quantum data bus for
transferring information. In a pioneering study~\cite{Bose}, Bose showed
that the simplest coupled spin chain with uniform nearest-neighbor (NN)
couplings is able to act as a quantum channel, i.e., the spin system allows
the transmission of an arbitrary quantum state with high fidelity from one
end to the other. The advantages of this protocol are that no external
control is required throughout the entire transfer process, the quantum
state transfer (QST) is equivalent to the natural dynamical evolution of the
time-independent Hamiltonian, and the system can be isolated from the
environment to minimize decoherence. However, the drawback of this proposal
is that the transfer quality decreases with the size of chain. One way to
overcome this problem is to precisely modulate the couplings between NN
spins throughout the quantum data bus, as suggested in Ref.~\cite{MC} so as
to obtain perfect QST, which is independent of the chain length. This is
possible because the eigenvalues of the system match the parity of the
corresponding eigenstates, which is a sufficient condition for perfect QST~%
\cite{Song,ST,LY2}. However, such an implementation requires precise control
of the system, which is not desirable in an experiment. Another approach to
achieving perfect QST is based on a gap quantum system~\cite%
{Cirac,LY1,HMX,XH,CB,Lukin,MB,AW0,AW,SP,SL,SWAP,LB,TJ,QI,Venuti3,Venuti2}.
By weakly connecting the transmitting and receiving qubits to a gap system,
the total system's dynamics can be reduced to those of an effective two- or
three-level system. In addition to the fact that no extra controls are
required for communication, a key advantage of these methods is their
robustness against parameter disorder, which comes from inevitable
technological errors in the experimental implementation. Moreover, we notice
that the systems with long-range inter-qubit interactions for perfect QST or
creating entanglement are well developed as well \cite%
{Venuti1,Petrosyan,YS,Kay,Bose3}.

In this paper, we introduce an impurity-induced gapped system (IGS), which
is a tight-binding chain with on-site energy applied on a single quantum dot
(QD), to act as a quantum channel. We demonstrate the existence of a
nonvanishing energy gap between the ground and first-excited states in the
single-particle case. We also investigate the QST using the IGS. It is found
that at lower temperatures, the total Hamiltonian can be mapped to a \textit{%
three-level} effective Hamiltonian whose energy levels are equally spaced
and can be used to perform near-perfect QST. In the weak-coupling limit, the
coupling constant of the effective Hamiltonian has an inverse relationship
with the transfer distance. Moreover, we study the robustness of the state
transfer against the static imperfections of the couplings, as discussed in
Ref. \cite{AZ,Chiara}. The resulting distribution of the transfer fidelities
reveals that chains with boundary states are more resilient to
imperfections. This is reflected in more instances of high-fidelity transfer
through the spin chain. Compared with previously proposed schemes, the
advantage of our scheme is that it is simple and can be readily applied to
experiments.

This paper is organized as follows: In Section II, the model IGS is set up
and its spectrum is introduced. In Section III, our QST protocol is set up
and the effective Hamiltonian, $H_{\text{eff}}$, is deduced using
perturbation theory. The scheme for using the IGS to transfer a quantum
state is discussed in Section IV. Finally, conclusions of these
investigations are presented in Section V.

\section{MODEL OF QUANTUM COMMUNICATION}

\subsection{Quantum data bus}

\begin{figure}[tbp]
\center
\includegraphics[bb=81 384 403 510, width=8 cm, clip]{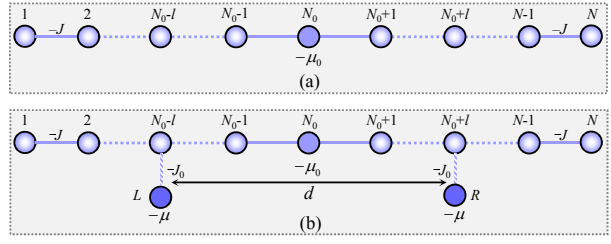}
\caption{(Color online) (a) Schematic illustrations of impurity-induced gap
system (IGS) which is a $N$-site chain with on-site energy applied on the
central quantum dot. (b) A schematic setup for QST between two QDs \textit{L}
and \textit{R} over the distance $d=2l+1$, through the gap system. }
\label{fig1}
\end{figure}

We begin by introducing a one-dimensional tight-binding chain of $N$ QDs with one diagonal impurity at $N_{0}$-th site, which acts as a
quantum data bus. The model is shown in Fig.~1(a), which is described by the
Hamiltonian%
\begin{equation}
\hat{H}_{M}^{e}=-J\sum_{j=1}^{N-1}\sum_{\sigma =\uparrow ,\downarrow }\left(
\hat{c}_{j,\sigma }^{\dag }\hat{c}_{j+1,\sigma }+\text{h.c.}\right) -\mu
_{0}\sum_{\sigma =\uparrow ,\downarrow }\hat{c}_{N_{0},\sigma }^{\dag }\hat{c%
}_{N_{0},\sigma },  \label{Hd}
\end{equation}%
where $-J$ $(<0)$ is the hopping amplitude between NN sites $j$ and $j+1$, $%
\hat{c}_{j,\sigma }^{\dag }$ and $\hat{c}_{j,\sigma }$ are the creation and
annihilation operators of electrons on the $j$-th site with spin $\sigma $,
and $-\mu _{0}$ $(<0)$ is the on-site energy of the defect. With a view
toward the quantum information, we can encode the qubit on the spin state.
Note that Eq.~(1) does not contain any spin-spin interaction term; thus, the
spin degree does not change during the evolution of the system. Hereafter,
we shall omit the $\sigma $ index, denoting the electron operator with
generic spin state as $\hat{a}_{j}^{\dag }=\cos \theta \hat{c}_{j,\uparrow
}^{\dagger }+e^{i\phi }\sin \theta \hat{c}_{j,\downarrow }^{\dagger }$. This
system can be regarded as a spinless fermion system, and the feasibly
obtained results can be applied to the original system. In this sense, we
can concentrate on the spinless fermion model in the following discussion.
\begin{equation}
\hat{H}_{M}=-J\sum_{j=1}^{N-1}\left( \hat{a}_{j}^{\dag }\hat{a}_{j+1}+\text{%
h.c.}\right) -\mu _{0}\hat{a}_{N_{0}}^{\dag }\hat{a}_{N_{0}},
\end{equation}%
For the sake of clarity and simplicity, we only consider the case where the
defect is placed in the middle of the medium, i.e., $N_{0}=(N+1)/2$. Note
that the Hamiltonian, $\hat{H}_{M}$, commutes with the total number
operator, $\hat{n}=\sum_{j=1}^{N}\hat{a}_{j}^{\dag }\hat{a}_{j}$, and so the
Hilbert space can be decomposed into subspaces corresponding to different
particle numbers, $n$. For the case of transferring a single particle, we
restrict the discussion to the single-particle subspace, which is spanned by
the Fock states $\left\vert j\right\rangle =\hat{a}_{j}^{\dag }\left\vert
0\right\rangle $, with $j=1,2,...,N$.

In this study, we focus on the bound state (or the ground state of $\hat{H}%
_{M}$) of this Hamiltonian for nonzero $\mu _{0}$, which can be obtained via
the Bethe ansatz method. We will also show that for Hamiltonian $\hat{H}_{M}$%
, there exists a finite energy gap, $\Delta =\varepsilon _{1}-\varepsilon
_{g}\sim \mu _{0}^{2}/2J$, between the ground state and the first excited
state.

To deduce the above conclusion, we write the state in the single-particle
Hilbert space as $\left\vert \lambda _{n}\right\rangle
=\sum_{j=1}^{N}f_{j}^{n}\left\vert j\right\rangle $. Substituting the
discrete superposition state into the eigenequation $\hat{H}_{M}\left\vert
\lambda _{n}\right\rangle =\varepsilon _{n}\left\vert \lambda
_{n}\right\rangle $, we get
\begin{equation}
-J\sum_{i=1}^{N}\left( \delta _{i,j-1}+\delta _{i,j+1}\right)
f_{i}^{n}=\left( \mu _{0}\delta _{j,N_{0}}+\varepsilon _{n}\right) f_{j}^{n},
\label{BA}
\end{equation}%
with open boundary condition $f_{0}^{n}=f_{N+1}^{n}=0$.

For $\mu _{0}=0$, the solution of Eq. (\ref{BA}) is
\begin{equation}
f_{j}^{n}=\sqrt{\frac{1}{N_{0}}}\sin \left[ \frac{(n+1)\pi j}{2N_{0}}\right]
,  \label{m0}
\end{equation}
with $n=0,$ $1,$ $2,\ldots ,$ $N-1$, the eigenvalues are $\varepsilon
_{n}=-2J\cos \left[ (n+1)\pi /2N_{0}\right] $.

We now study the effect of the impurity on the energy spectrum of
Hamiltonian $\hat{H}_{M}$\ for nonzero $\mu _{0}$. Before making
calculations, we make the following observations: first, when the
Hamiltonian, $\hat{H}_{M}$, is processing mirror symmetry with respect to
the inversion center, $N_{0}$, its eigenvectors, $\left\vert \lambda
_{n}\right\rangle $, have definite parities. Moreover, if the eigenvalues, $%
\lambda _{n}$, are in increasing order, then the eigenvectors, $\left\vert
\lambda _{n}\right\rangle $, change parity alternatively, i.e., the
mirror-inverted eigenstates, $\left\vert \lambda _{n}\right\rangle $,
satisfy the relation $f_{j}^{n}=(-1)^{n}f_{N+1-j}^{n}$ upon assuming that
even (odd) $n$ label even (odd) eigenstates $\left\vert \lambda
_{n}\right\rangle $. Second, the probability density of all the eigenstates
with odd parity in the central site, $N_{0}$, is zero, i.e., $%
f_{N_{0}}^{2m-1}=0$, which means that the eigenstates with odd parity are
unaffected by the presence of the impurity. Third, by the Hellmann-Feynman
theorem, the eigenvalues of even-parity eigenstates decrease due to the
presence of the impurity. Furthermore, the impurity contributes exactly one
bound state, which we focus on in this study.

To see more precisely what happens for $\mu _{0}\neq 0$, we solve Eq. (\ref%
{BA}) via the Bethe Ansatz method. In this study, the bound state is the
ground state of $\hat{H}_{M}$. Through a straightforward calculation, one
can obtain the following analytical result for the ground state
\begin{equation}
f_{j}^{0}=\Omega ^{-1/2}\times \left\{
\begin{array}{c}
\sinh k_{0}j,\text{ \ \ \ \ \ \ \ \ \ \ \ \ \ \ }j\leq N_{0} \\
\sinh k_{0}(N+1-j),\text{ }j>N_{0}%
\end{array}%
\right. ,
\end{equation}%
with the eigenvalue $\lambda _{0}=-2J\sqrt{\xi ^{2}+1}$, where $k_{0}=\ln %
\left[ \xi +\sqrt{\xi ^{2}+1}\right] $ and $\xi =\mu _{0}/2J$; $\Omega
\approx e^{2k_{0}N_{0}}\left( e^{2k_{0}}+1\right) /4\left(
e^{2k_{0}}-1\right) $ is the renormalization factor.

The remaining eigenstates with \textit{even parity} are extended and similar
to Eq. (\ref{m0}); the appropriate Ansatz is
\begin{equation}
f_{j}^{n}=\left\{
\begin{array}{c}
\sin k_{n}j,\text{ \ \ \ \ \ \ \ \ \ \ \ \ \ \ }j\leq N_{0} \\
\sin k_{n}(N+1-j),\text{ }j>N_{0}%
\end{array}%
\right. ,
\end{equation}%
which yields the eigenvalue $\varepsilon _{n}=-2J\cos k_{n}$ and the wave
vector, $k_{n}$, obeys
\begin{equation}
\xi \sin k_{n}N_{0}=\cos k_{n}N_{0}\sin k_{n}.  \label{wv}
\end{equation}%
Setting $\tan \varphi _{n}=\xi /\sin k_{n}$, Eq. (\ref{wv}) becomes $\cos
\left( k_{n}N_{0}+\varphi _{n}\right) =0$, whose allowed values are
\begin{equation}
k_{n}=\frac{\left( 2m-1\right) \pi -2\varphi _{n}}{2N_{0}},m=1,2...,N_{0}-1.
\end{equation}%
From the above equations, we know that (i) the phase shift $\varphi _{n}=0$
for $\xi =0$ and $\varphi _{n}=\pi /2$ for $\xi =\infty $, and that (ii) the
phase shifts do not alter the order of the sequence $\left\{ k_{n}\right\} $.

\begin{figure}[tbp]
\center
\includegraphics[bb=86 339 430 574, width=7 cm, clip]{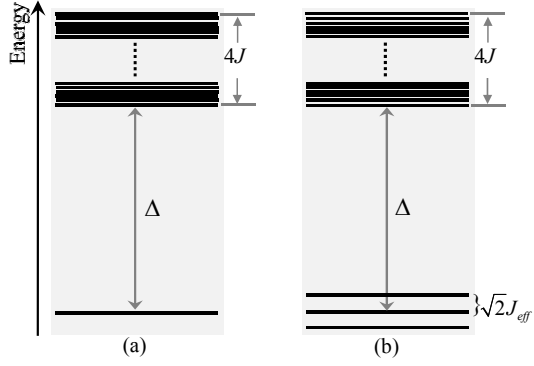}
\caption{Schematic illustration of the energy levels of the system. (a) When
the connections between two QDs and the medium switch off $J_{0}=0$ the
ground states are three-fold degenerate. (b) When $J_{0}=0$ switches on, the
degeneracy will be removed and split into three sub-levels with the level
spacing $\protect\delta =\protect\sqrt{2}J_{\text{eff}}$. This is
approximately equivalent to that of tight-binding chain with three QDs.}
\label{fig2}
\end{figure}

Until now, we have only discussed the solutions of eigenequation $\hat{H}%
_{M}\left\vert \lambda _{n}\right\rangle =\varepsilon _{n}\left\vert \lambda
_{n}\right\rangle $ without any external perturbation. In the thermodynamic
limit where $N_{0}\rightarrow \infty $, the excited energies become a
continuous energy band; it is not hard to find that the energy gap between
the ground state and the first excited state (see the Fig.~\ref{fig2}a) is
\begin{equation}
\Delta =2J\sqrt{\xi ^{2}+1}-2J.
\end{equation}%
For very small values of onsite energy, i.e., $\mu _{0}\ll J$, we get $%
\Delta \approx J\xi ^{2}$.

\subsection{The subspace Hamiltonian}

Now, let us introduce the protocol of quantum communication by using IGS, in
which two individual QDs (sender and receiver) are symmetrically coupled to
an IGS on opposite sides of the data bus (as illustrated in Fig. 1(b)).
Moreover, QDs \textit{L} and \textit{R} are supplied with on-site energy, $%
-\mu $. The total Hamiltonian consisting of $(N+2)$ QDs reads
\begin{eqnarray}
\hat{H} &=&\hat{H}_{0}+\hat{H}^{\prime },  \notag \\
\hat{H}_{0} &=&\hat{H}_{M}-\mu \left( \hat{a}_{L}^{\dag }\hat{a}_{L}+\hat{a}%
_{R}^{\dag }\hat{a}_{R}\right) ,  \notag \\
\hat{H}^{\prime } &=&-J_{0}\left( \hat{a}_{L}^{\dagger }\hat{a}_{N_{0}-l}+%
\hat{a}_{R}^{\dagger }\hat{a}_{N_{0}+l}+\text{h.c.}\right) ,  \label{H}
\end{eqnarray}%
where $\hat{a}_{L}$ and $\hat{a}_{R}$ are the annihilation operators of
electron on $L$ and $R$, $\left( N_{0}\pm l\right) $ denotes the connecting
sites of the chain, and the coupling constant, $J_{0}$, measures the
strength of the interaction.

In the absence of coupling between the two qubits and the medium ($J_{0}=0$)
and setting $-\mu =\lambda _{0}$, the total Hamiltonian (\ref{H}) can be
diagonalized in the basis $\left\{ \left\vert L\right\rangle ,\left\vert
R\right\rangle ,\left\vert \lambda _{0}\right\rangle ,\left\vert \lambda
_{1}\right\rangle ,\ldots ,\left\vert \lambda _{N-1}\right\rangle \right\} $%
, and its ground states are threefold degenerate, i.e., $\left\vert
L\right\rangle $, $\left\vert \lambda _{0}\right\rangle $, and $\left\vert
R\right\rangle $ have the energy $E_{g}^{\left( 0\right) }=-\mu $. These
three states can be regarded as the components of an effective pseudo-spin-1
system that span an invariant subspace. The original energy degeneracy will
break down by switching on the weak coupling, $J_{0}$ ($J_{0}\ll \Delta $),
and the ground state will split into three sub-levels with level spacing $%
\Delta E=\sqrt{2}\left\vert J_{\text{eff}}\right\vert $, as illustrated in
Fig.~\ref{fig2}(b). Here, $J_{\text{eff}}$ is the effective hopping integral
that can be calculated as follows.

When switching on $J_{0}$, the eigenequation becomes $(\hat{H}_{0}+\hat{H}%
^{\prime })\left\vert \psi \right\rangle =E\left\vert \psi \right\rangle $.
For weak coupling between QDs and the bus, $J_{0}\ll \Delta $, $\hat{H}%
^{\prime }$ can be treated as a perturbation Hamiltonian. Let us assume
that, in some definite way, we can divide the basis into two classes, $\left[
G\right] =\left\{ \left\vert L\right\rangle ,\left\vert R\right\rangle
,\left\vert \lambda _{0}\right\rangle \right\} $ where the relative
complement of $\left[ G\right] $ is denoted by $\left[ O\right] =\left\{
\left\vert \lambda _{1}\right\rangle ,\ldots ,\left\vert \lambda
_{N-1}\right\rangle \right\} $. Defining
\begin{eqnarray}
\mathcal{\hat{G}} &=&\left\vert L\right\rangle \left\langle L\right\vert
+\left\vert \lambda _{0}\right\rangle \left\langle \lambda _{0}\right\vert
+\left\vert R\right\rangle \left\langle R\right\vert , \\
\mathcal{\hat{O}} &=&\sum_{n=1}^{N-1}\left\vert \lambda _{n}\right\rangle
\left\langle \lambda _{n}\right\vert ,
\end{eqnarray}%
denote two orthogonal projection operators of two different subspaces. It is
easy to check that $\mathcal{\hat{G}\hat{O}}=0$ and satisfying $\mathcal{%
\hat{G}}+\mathcal{\hat{O}}=\mathbb{\hat{I}}$. The eigenequation can be
rewritten as%
\begin{equation}
\left( \mathcal{\hat{G}}+\mathcal{\hat{O}}\right) \hat{H}\left( \mathcal{%
\hat{G}}+\mathcal{\hat{O}}\right) \left( \mathcal{\hat{G}}+\mathcal{\hat{O}}%
\right) \left\vert \psi \right\rangle =E\left( \mathcal{\hat{G}}+\mathcal{%
\hat{O}}\right) \left\vert \psi \right\rangle .
\end{equation}%
The above equation can be decomposed into two basic formulae in subspaces $%
\left( G\right) $ and $\left( O\right) $
\begin{eqnarray}
\hat{H}_{\mathcal{GG}}\mathcal{\hat{G}}\left\vert \psi \right\rangle +\hat{H}%
_{\mathcal{GO}}\mathcal{\hat{O}}\left\vert \psi \right\rangle &=&E\mathcal{%
\hat{G}}\left\vert \psi \right\rangle ,  \label{PPsi} \\
\hat{H}_{\mathcal{OG}}\mathcal{\hat{G}}\left\vert \psi \right\rangle +\hat{H}%
_{\mathcal{OO}}\mathcal{\hat{O}}\left\vert \psi \right\rangle &=&E\mathcal{%
\hat{O}}\left\vert \psi \right\rangle ,  \label{OPsi}
\end{eqnarray}%
where $\hat{H}_{\alpha \beta }=\hat{\alpha}\hat{H}\hat{\beta}$, $\left( \hat{%
\alpha},\hat{\beta}=\mathcal{\hat{G}},\mathcal{\hat{O}}\right) $. Using Eq.~ %
\eqref{OPsi}, one can express $\mathcal{\hat{O}}\left\vert \psi
\right\rangle $ in terms of $\mathcal{\hat{G}}\left\vert \psi \right\rangle $%
:
\begin{equation}
\mathcal{\hat{O}}\left\vert \psi \right\rangle =\left( E-\hat{H}_{\mathcal{OO%
}}\right) ^{-1}\hat{H}_{\mathcal{OG}}\mathcal{\hat{G}}\left\vert \psi
\right\rangle ,
\end{equation}%
so that, substituting the above equation into Eq. \eqref{PPsi}, one finds
that, to second order, the equation only evolves $\mathcal{\hat{G}}%
\left\vert \psi \right\rangle $:%
\begin{equation}
\hat{H}_{\text{eff}}\mathcal{\hat{G}}\left\vert \psi \right\rangle =E%
\mathcal{\hat{G}}\left\vert \psi \right\rangle ,  \label{Heff}
\end{equation}%
where
\begin{equation}
\hat{H}_{\text{eff}}=\hat{H}_{\mathcal{GP}}+\hat{H}_{\mathcal{PO}}\left( E-%
\hat{H}_{\mathcal{OO}}\right) ^{-1}\hat{H}_{\mathcal{OG}}
\end{equation}%
denotes the effectvie Hamiltonian in subspace $\left( A\right) $ with
\begin{eqnarray}
\hat{H}_{\mathcal{GG}} &=&-J_{0}\zeta _{0}\left( \left\vert L\right\rangle
+\left\vert R\right\rangle \right) \left\langle \lambda _{0}\right\vert +%
\text{h.c.}-\mu \mathcal{\hat{G}},  \label{HPP} \\
\hat{H}_{\mathcal{OO}} &=&\sum_{n=1}^{N-1}\lambda _{n}\left\vert \lambda
_{n}\right\rangle \left\langle \lambda _{n}\right\vert ,  \label{HOO} \\
\hat{H}_{\mathcal{GO}} &=&\sum_{n=1}^{N-1}-J_{0}\zeta _{n}\left( l\right) %
\left[ \left\vert L\right\rangle +(-1)^{n}\left\vert R\right\rangle \right]
\left\langle \lambda _{n}\right\vert .  \label{HPO}
\end{eqnarray}%
and $\zeta _{n}\left( l\right) =\langle N_{0}-l\left\vert \lambda
_{n}\right\rangle $. Through a straightforward calculation, one can obtain%
\begin{equation*}
\hat{H}_{\mathcal{GO}}\left( \mu +\hat{H}_{\mathcal{OO}}\right) ^{-1}\hat{H}%
_{\mathcal{OG}}=\sum_{n=1}^{N-1}\frac{J_{0}^{2}\left\vert \zeta
_{n}\right\vert ^{2}}{E-\lambda _{n}}\left[ \left\vert L\right\rangle
+\left( -1\right) ^{n}\left\vert R\right\rangle \right] \left[ \left\langle
L\right\vert +\left( -1\right) ^{n}\left\langle R\right\vert \right] .
\end{equation*}

Note that the eigenvalues, $E$, determined from Eq. (\ref{Heff}), are
perturbed eigenvalues around respective unperturbed value $-\mu $. With this
connection, one seldom requires the second-order correction, which is small (%
$J_{0}^{2}\ll \left\vert E-\lambda _{n}\right\vert $, which is the condition
for the perturbation procedure to be a good approximation in this problem);
it is therefore sufficient to quote the first-order results%
\begin{equation}
\hat{H}_{\text{eff}}\approx -J_{\text{eff}}\left( \left\vert L\right\rangle
+\left\vert R\right\rangle \right) \left\langle \lambda _{0}\right\vert -%
\frac{\mu }{2}\left( \left\vert L\right\rangle \left\langle L\right\vert
+\left\vert \lambda _{0}\right\rangle \left\langle \lambda _{0}\right\vert
+\left\vert R\right\rangle \left\langle R\right\vert \right) +\text{h.c.},
\label{H3}
\end{equation}%
with effective coupling strength $J_{\text{eff}}=J_{0}\zeta _{0}\left(
l\right) $.

In this section, we have shown that the total Hamiltonian~(\ref{H}) can be
simplified to the effective Hamiltonian~(\ref{H3}), due to a large gap
(compared with coupling strength $J_{0}$) existing in the medium. This
approximation holds when the energy splitting, $\sqrt{2}J_{\text{eff}}$,
caused by the $\hat{H}_{\text{eff}}$ is smaller than the typical gap for the
unperturbed Hamiltonian, $\hat{H}_{0}$, i.e., $J_{\text{eff}}\ll \Delta $.
To check the range of validity of the above effective Hamiltonian, we
compare the analytical result of $J_{\text{eff}}$ with the results $\left(
E_{1}-E_{0}\right) /\sqrt{2}$ obtained by direct numerical diagonalization
of the Hamiltonian~(\ref{H}). The results of this comparison are plotted in
Fig. 3 for a system of $N=499$, with $J_{0}=0.001J$, and $\mu _{0}=0.1J$, $%
0.05J$, and $0.01J$. In this figure, one can see that taking $\mu
_{0}/J_{0}\ $bigger than $50$, the effective coupling strength, $J_{\text{eff%
}}$, of $\hat{H}_{\text{eff}}$\ agrees very well with that obtained
numerically. So far, the validity of the effective Hamiltonian~(\ref{H3}) is
firmly established. Thus one should be able to obtain high-fidelity QST with
the effective Hamiltonian whenever the perturbation solution is valid.
Furthermore, we will show that the existence of an energy gap can also be
used to protect the performance of QST in the presence of static disorder in
the couplings of the quantum data bus.

However, it is worth pointing out that large $\mu _{0}$ can improve the
validity of $\hat{H}_{\text{eff}}$ but decrease the transfer efficiency
characterized by $J_{\text{eff}}$, since $1/J_{\text{eff}}$ determines the
transfer time of the QST between the two qubits, \textit{L} and \textit{R}.
As observed in Fig. 3, the decay rate of $J_{\text{eff}}$ directly depends
on the value of $\mu _{0}$. The smaller the $\mu _{0}$ is, the slower the
decay rate will be. Typically, $J_{\text{eff}}$ decreases almost linearly
with the increase of the transfer distance for $\mu _{0}=0.01J$. From the
two competing aspects described above, we can summarize the proper choice of
the system parameters, $\mu _{0}$ and $J_{0}$, for high-fidelity QST.

\begin{figure}[tbp]
\center
\includegraphics[bb=79 471 385 710, width=7 cm, clip]{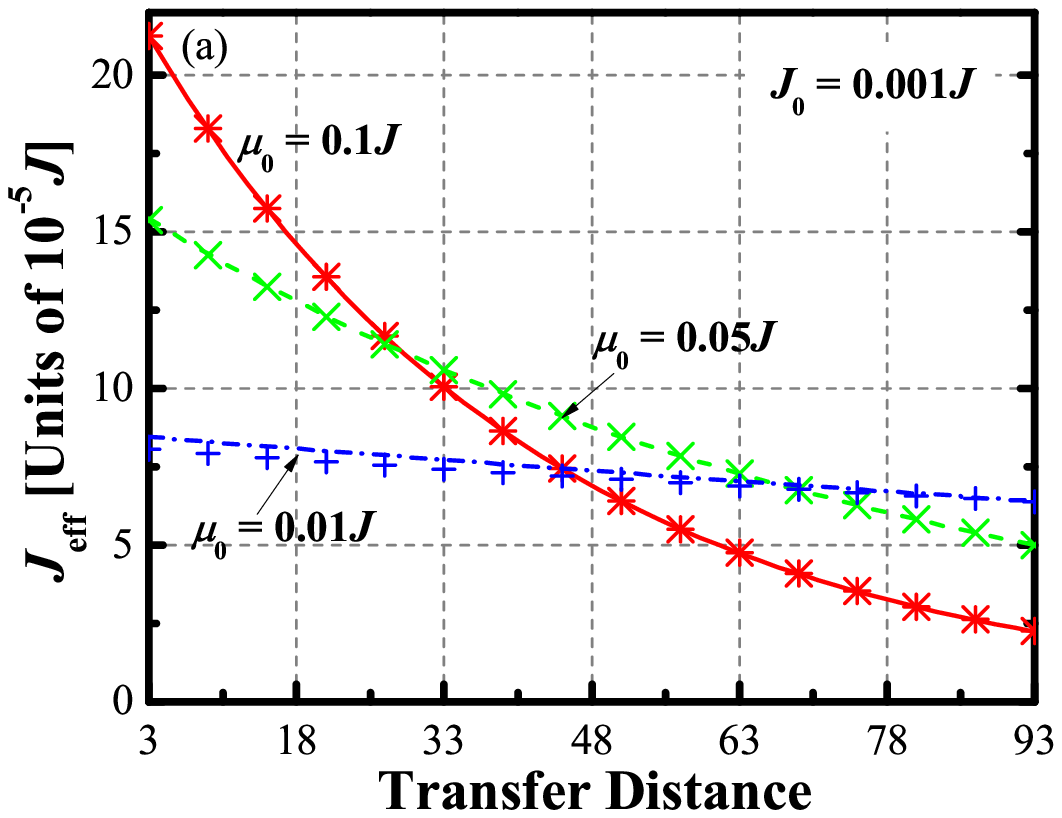} %
\includegraphics[bb=79 471 385 710, width=7 cm, clip]{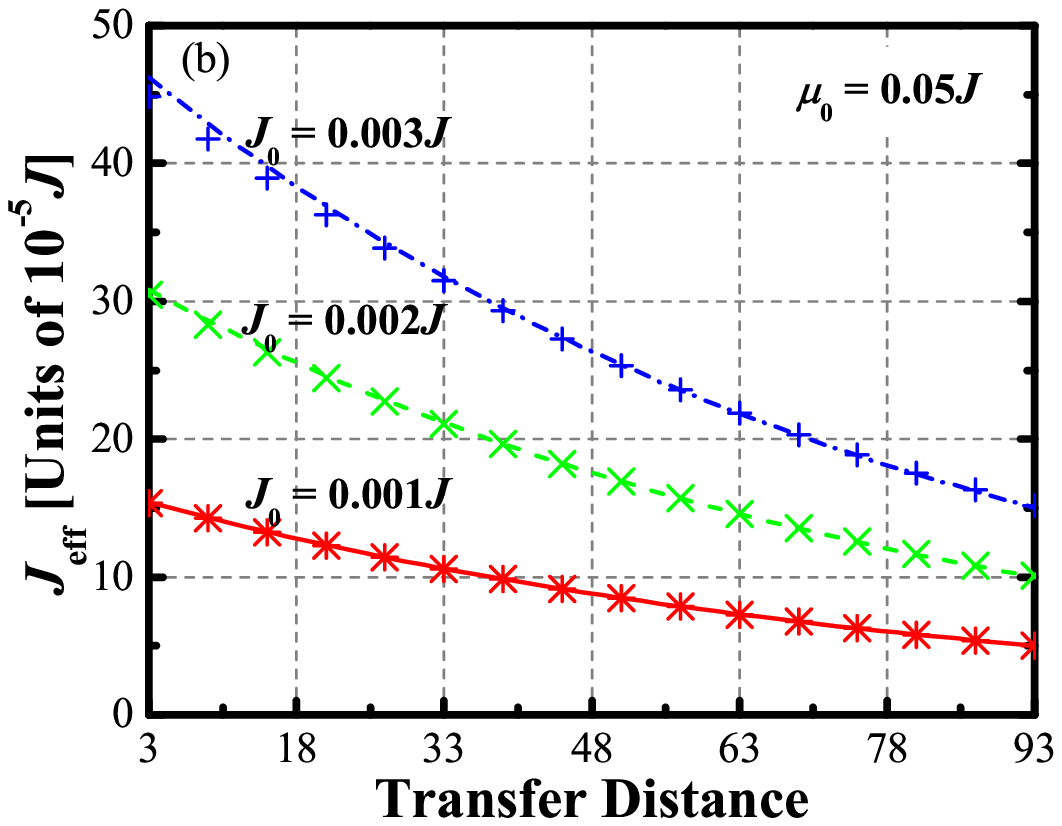}
\caption{(Color online) (a) Behavior of $J_{\text{eff}}$ as a function of
the transfer distance $d$ for a chain of $N=499$ sites, $J_{0}=0.001$ and
different values of $\protect\mu _{0}$. Curves from top to bottom are,
respectively, for $\protect\mu _{0}=0.1J$, $0.05J$, and $0.01J$. Continuous
curves display the approximate expression for the $J_{\text{eff}}$ and
symbols denote the exact numerical data which is given by $\left(
E_{1}-E_{g}\right) /\protect\sqrt{2}$. (b) The same as in (a) but for $%
\protect\mu _{0}=0.05J$ and different values of $J_{0}$.}
\label{fig3}
\end{figure}

To briefly summarize, we have theoretically and numerically studied $J_{%
\text{eff}}$ as a function of $d$ in a specific range of parameters.
However, the obtained conclusion is based on the fact that the $\hat{H}_{%
\text{eff}}$ given by Eq.~(\ref{H3}) is a valid approximation in the studied
range. In the following discussion, the validity of $\hat{H}_{\text{eff}}$
is investigated by comparing the eigenstates of $\hat{H}_{\text{eff}}$ with
the lowest three states of the total system~(\ref{H}).

Define the quasi-angular momentum states $\left\vert j,m\right\rangle $ as
\begin{eqnarray}
\left\vert 1,0\right\rangle &=&\frac{1}{\sqrt{2}}\left( \left\vert
L\right\rangle -\left\vert R\right\rangle \right) , \\
\left\vert 1,\pm 1\right\rangle &=&\frac{1}{2}\left( \left\vert
L\right\rangle \pm \sqrt{2}\left\vert \lambda _{0}\right\rangle +\left\vert
R\right\rangle \right) ,
\end{eqnarray}%
which are the eigenstates of effective Hamiltonian~(\ref{H3}). On the other
hand, the eigenstates of $\hat{H}$ can be generally written as%
\begin{equation}
\left\vert \psi _{jm}\right\rangle =c_{L}\left\vert L\right\rangle
+\sum_{n=0}^{N-1}c_{n}\left\vert \lambda _{n}\right\rangle +c_{R}\left\vert
R\right\rangle ,
\end{equation}%
where we have the condition $\left\vert c_{L}\right\vert
^{2}+\sum_{n}\left\vert c_{n}\right\vert ^{2}+\left\vert c_{R}\right\vert
^{2}=1$ for the normalization of $\left\vert \psi _{jm}\right\rangle $.
Moreover, we assign the state $\left\vert \psi _{jm}\right\rangle $ to
denote the ground state for $j=1$, $m=1$, the first excited state for $j=1$,
$m=0$, and the second excited state for $j=1$, $m=-1$. To evaluate the
fidelity of the $\hat{H}_{\text{eff}}$ induced by the perturbation, we
introduce the overlap
\begin{equation}
P_{jm}=\left\vert \left\langle j,m\right. \left\vert \psi _{jm}\right\rangle
\right\vert ^{2}.
\end{equation}

For the case where $J_{0}=0$, the ground states $\left\vert \psi
_{jm}\right\rangle $ of $\hat{H}$ are threefold degenerate and $\left\vert
\psi _{jm}\right\rangle $ can be written in symmetrical form by linear
combinations of $\left\vert L\right\rangle $, $\left\vert \lambda
_{0}\right\rangle $, and $\left\vert R\right\rangle $. Under this condition,
one can obtain $P_{jm}=1$ for $j=1$ and $m=0$, $\pm 1$. In particular, we
have $\left\vert c_{L}\right\vert ^{2}=\left\vert c_{R}\right\vert ^{2}=1/4$%
, $\left\vert c_{0}\right\vert ^{2}=1/2$ for $m=\pm 1$ and $\left\vert
c_{L}\right\vert ^{2}=\left\vert c_{R}\right\vert ^{2}=1/2$, $\left\vert
c_{0}\right\vert ^{2}=0$ for $m=0$. For the practical Hamiltonian $\hat{H}$,
i.e., $J_{0}\neq 0$, the values of $\left\vert c_{i}\right\vert ^{2}$ $%
\left( i=L,0,R\right) $ and $P_{jm}$ are numerically calculated for the
three lowest eigenstates in the $N=499$ system with $\mu _{0}=0.1J$, $0.05J$
and $J_{0}=2\times 10^{-3}J$ for finite transfer distances $d=5$, , $15$, $%
25 $, $35$, $45$, $55$, and $65$, which are listed in Tables I(a) and (b).

\begin{table*}[tbp]
\caption{The overlap $P_{jm}$ and its three components, which provide a
criteria for the validity of $H_{\text{eff}}$, are calculated numerically
for the ground state, first excited state and second excited state of total
system for finite transfer distance $d=5$, $15$, $25$, $35$, $45$, $55$, and
$65$. The results for $\protect\mu _{0}=0.1J$, and $0.05J$ ($J_{0}=2\times
10^{-3}J$) are listed in (a), and (b) respectively. It shows that the result
based on the realistic interaction is very close to that by $H_{\text{eff}}$
even if $\protect\mu _{0}$ is not large enough.}
\begin{center}
$%
\begin{tabular}{ccccccccccc}
\hline\hline
States & $\ j\ $ & $m$ & $d=$ & \ \ 5\ \ \  & \ \ \ \ 15\ \ \ \  & \ \ \ 25\
\ \  & \ \ \ \ 35\ \ \ \  & \ \ \ \ 45\ \ \ \  & \ \ \ 55\ \ \  & \ \ \ 65\
\ \  \\ \hline
(a) &  &  & $\left\vert c_{L}\right\vert ^{2}$ & 0.2552 & 0.2531 & 0.2539 &
0.2569 & 0.2618 & 0.2687 & 0.2778 \\
$\left\vert \psi _{11}\right\rangle $ & $1$ & $1$ & $\left\vert
c_{0}\right\vert ^{2}$ & 0.4884 & 0.4932 & 0.4920 & 0.4860 & 0.4757 & 0.4613
& 0.4426 \\
&  &  & $\left\vert c_{R}\right\vert ^{2}$ & 0.2552 & 0.2531 & 0.2539 &
0.2569 & 0.2618 & 0.2687 & 0.2778 \\
&  &  & $P_{11}$ & 0.9986 & 0.9994 & 0.9997 & 0.9995 & 0.9988 & 0.9973 &
0.9950 \\
&  &  & $\left\vert c_{L}\right\vert ^{2}$ & 0.4999 & 0.4994 & 0.4987 &
0.4980 & 0.4975 & 0.4971 & 0.4968 \\
$\left\vert \psi _{10}\right\rangle $ & $1$ & $0$ & $\left\vert
c_{0}\right\vert ^{2}$ & 3.457$\times $10$^{-25}$ & 2.334$\times $10$^{-24}$
& 6.501$\times $10$^{-25}$ & 2.248$\times $10$^{-23}$ & 1.044$\times $10$%
^{-23}$ & 2.920$\times $10$^{-23}$ & 1.992$\times $10$^{-22}$ \\
&  &  & $\left\vert c_{R}\right\vert ^{2}$ & 0.4999 & 0.4994 & 0.4987 &
0.4980 & 0.4975 & 0.4971 & 0.4968 \\
&  &  & $P_{10}$ & 0.9999 & 0.9988 & 0.9973 & 0.9960 & 0.9949 & 0.9941 &
0.9935 \\
&  &  & $\left\vert c_{L}\right\vert ^{2}$ & 0.2432 & 0.2462 & 0.2459 &
0.2429 & 0.2375 & 0.2301 & 0.2205 \\
$\left\vert \psi _{1-1}\right\rangle $ & $1$ & $-1$ & $\left\vert
c_{0}\right\vert ^{2}$ & 0.5116 & 0.5068 & 0.5080 & 0.5140 & 0.5243 & 0.5387
& 0.5574 \\
&  &  & $\left\vert c_{R}\right\vert ^{2}$ & 0.2432 & 0.2462 & 0.2459 &
0.2429 & 0.2375 & 0.2301 & 0.2205 \\
&  &  & $P_{1-1}$ & 0.9979 & 0.9992 & 0.9997 & 0.9995 & 0.9987 & 0.9973 &
0.9950 \\
&  &  &  &  &  &  &  &  &  &  \\
(b) &  &  & $\left\vert c_{L}\right\vert ^{2}$ & 0.2641 & 0.2599 & 0.2580 &
0.2580 & 0.2595 & 0.2625 & 0.2666 \\
$\left\vert \psi _{11}\right\rangle $ & $1$ & $1$ & $\left\vert
c_{0}\right\vert ^{2}$ & 0.4633 & 0.4741 & 0.4801 & 0.4817 & 0.4795 & 0.4738
& 0.4650 \\
&  &  & $\left\vert c_{R}\right\vert ^{2}$ & 0.2641 & 0.2599 & 0.2580 &
0.2580 & 0.2595 & 0.2625 & 0.2666 \\
&  &  & $P_{11}$ & 0.9904 & 0.9933 & 0.9957 & 0.9973 & 0.9981 & 0.9980 &
0.9970 \\
&  &  & $\left\vert c_{L}\right\vert ^{2}$ & 0.4999 & 0.4985 & 0.4963 &
0.4937 & 0.4912 & 0.4887 & 0.4865 \\
$\left\vert \psi _{10}\right\rangle $ & $1$ & $0$ & $\left\vert
c_{0}\right\vert ^{2}$ & 6.467$\times $10$^{-24}$ & 2.461$\times $10$^{-25}$
& 1.581$\times $10$^{-23}$ & 2.075$\times $10$^{-23}$ & 2.362$\times $10$%
^{-23}$ & 7.620$\times $10$^{-24}$ & 2.793$\times $10$^{-24}$ \\
&  &  & $\left\vert c_{R}\right\vert ^{2}$ & 0.4999 & 0.4985 & 0.4963 &
0.4937 & 0.4912 & 0.4887 & 0.4865 \\
&  &  & $P_{10}$ & 0.9997 & 0.9970 & 0.9925 & 0.9875 & 0.9823 & 0.9774 &
0.9730 \\
&  &  & $\left\vert c_{L}\right\vert ^{2}$ & 0.2194 & 0.2286 & 0.2348 &
0.2379 & 0.2383 & 0.2360 & 0.2315 \\
$\left\vert \psi _{1-1}\right\rangle $ & $1$ & $-1$ & $\left\vert
c_{0}\right\vert ^{2}$ & 0.5318 & 0.5231 & 0.5185 & 0.5177 & 0.5203 & 0.5261
& 0.5350 \\
&  &  & $\left\vert c_{R}\right\vert ^{2}$ & 0.2194 & 0.2286 & 0.2348 &
0.2379 & 0.2383 & 0.2360 & 0.2315 \\
&  &  & $P_{1-1}$ & 0.9684 & 0.9791 & 0.9874 & 0.9931 & 0.9963 & 0.9974 &
0.9967 \\ \hline\hline
\end{tabular}%
$%
\end{center}
\end{table*}

We remark that the condition for mapping $\hat{H}$ to the effective
Hamiltonian~(\ref{H3}) is that $J_{0}$ must be small enough compared to the
energy gap $\Delta $ of the medium rather than the on-site energy $\mu _{0}$%
. As mentioned before, the energy gap is $\Delta \approx \mu _{0}^{2}/4J$
for small $\mu _{0}$ (compared with $J$). It is straightforward to obtain $%
\Delta \approx 2.5\times 10^{-3}J$ for $\mu _{0}=0.1J$ and $\Delta \approx
6.25\times 10^{-4}J$ for $\mu _{0}=0.05J$. From the numerical results shown
in Table I, we observe that the realistic interaction leads to the results
for $\left\vert c_{i}\right\vert ^{2}$ $\left( i=L,0,R\right) $, which are
very close to those described by $\hat{H}_{\text{eff}}$, even if $\Delta $
is of the same order of $J_{0}$. It is clear that such a three-level
subsystem allows state $\left\vert L\right\rangle $ to transfer with high
fidelity, and the coherent population exhibits oscillations between the QDs
on the two ends. The oscillation period of the population is given by $\tau
=\pi /\sqrt{2}J_{\text{eff}}$, and we can say that the quantum state is
transferred from QD $L$ to QD $R$ at the time $\tau =\left( 2n-1\right)
\times \tau $.

\section{QUANTUM STATE TRANSFER}

\subsection{Weak Coupling Regime}

Note that the spectrum structure and the corresponding parity of the
effective Hamiltonian, $\hat{H}_{\text{eff}}$, obey the spectrum-parity
matching condition (SPMC)~\cite{ST,LY2} exactly, which is the general
criterion for perfect QST. In this section, we consider the QST scheme based
on our system. Assume Alice is at the sender site, A, and Bob is at the
receiver site, B. Let Alice hold an electron with a spin state that she
wants to communicate to Bob of $\left\vert \varphi \right\rangle =\cos
\left( \theta /2\right) \left\vert \uparrow \right\rangle +e^{i\phi }\sin
\left( \theta /2\right) \left\vert \downarrow \right\rangle $, where $%
\left\vert \uparrow \right\rangle $ $\left( \left\vert \downarrow
\right\rangle \right) $ denotes the spin-up (down) state. Thus, the initial
state of the total system is $\left\vert \Psi \left( 0\right) \right\rangle
=\left\vert L\right\rangle =\left( \cos \theta c_{L,\uparrow }^{\dagger
}+e^{i\phi }\sin \theta c_{L,\downarrow }^{\dagger }\right) \left\vert
0\right\rangle _{L}$, which is a superposition of the eigenstates of
Hamiltonian $\hat{H}_{\text{eff}}$
\begin{equation}
\left\vert \Psi \left( 0\right) \right\rangle =\frac{1}{2}\left( \left\vert
1,1\right\rangle +\sqrt{2}\left\vert 1,0\right\rangle +\left\vert
1,-1\right\rangle \right) .
\end{equation}%
At time $t$, the initial state $\left\vert \Psi \left( 0\right)
\right\rangle $\ evolves into%
\begin{eqnarray}
\left\vert \Psi \left( t\right) \right\rangle &=&e^{-i\hat{H}_{\text{eff}%
}t}\left\vert \Psi \left( 0\right) \right\rangle  \notag \\
&=&\frac{1}{2}\left( e^{i\delta t}\left\vert 1,1\right\rangle +\sqrt{2}%
\left\vert 1,0\right\rangle +e^{-i\delta t}\left\vert 1,-1\right\rangle
\right)
\end{eqnarray}%
where $\delta =\sqrt{2}J_{\text{eff}}$, and we have neglected the overall
phase, $e^{-i\varepsilon _{g}^{(0)}t}$. The density matrix corresponding to $%
\left\vert \Psi \left( t\right) \right\rangle $ is $\rho =\left\vert \Psi
\left( t\right) \right\rangle \left\langle \Psi \left( t\right) \right\vert $%
, and the probability of state $\left\vert \Psi \left( 0\right)
\right\rangle $ transferring to the QD \textit{R }at time\textit{\ }$t$ is
defined as%
\begin{equation}
F(t)=\text{Tr}(\rho \rho _{R})=\sin ^{4}\left( \frac{\delta t}{2}\right) .
\end{equation}%
At the moment when $t=\tau =\pi /\delta $, $F(\tau )=1$ indicates that our
scheme can perform QST perfectly. That is to say, the system evolves into a
new factorized state%
\begin{eqnarray}
\left\vert \Psi \left( \tau \right) \right\rangle &=&\frac{1}{2}\left(
e^{i\pi }\left\vert 1,1\right\rangle +\sqrt{2}\left\vert 1,0\right\rangle
+e^{-i\pi }\left\vert 1,-1\right\rangle \right)  \notag \\
&=&e^{i\pi }\left\vert R\right\rangle .
\end{eqnarray}

As an example of verifying the validity of the effective Hamiltonian $\hat{H}%
_{\text{eff}}$, the fidelity for $N=499$ and transfer distance $d=5$, with $%
J_{0}=2\times 10^{-3}J$, and $\mu _{0}=0.1J$, $0.05J$ are plotted in Figs.
4(a) and (b). They show that small $J_{0}$ leads to a result for transfer
fidelity, which is very close to that described by the effective
Hamiltonian, $\hat{H}_{\text{eff}}$.

\begin{figure}[tbp]
\center
\includegraphics[bb=82 469 379 707, width=7 cm, clip]{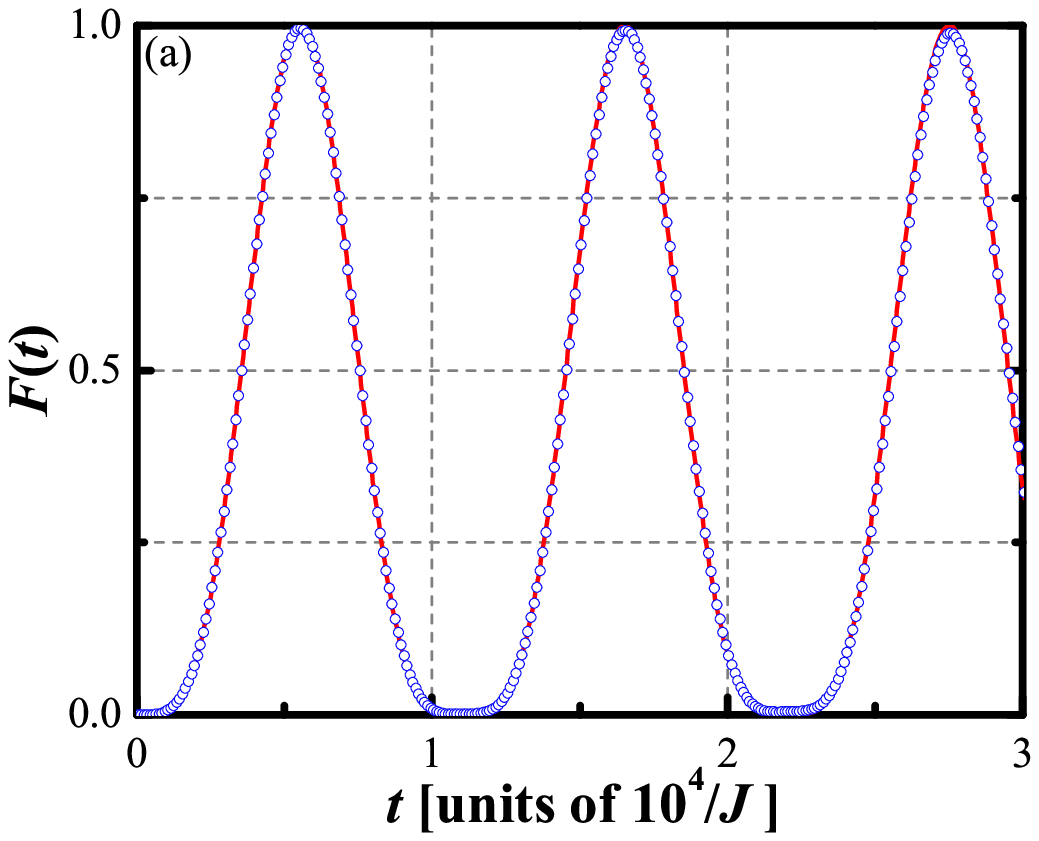} %
\includegraphics[bb=82 469 379 707, width=7 cm, clip]{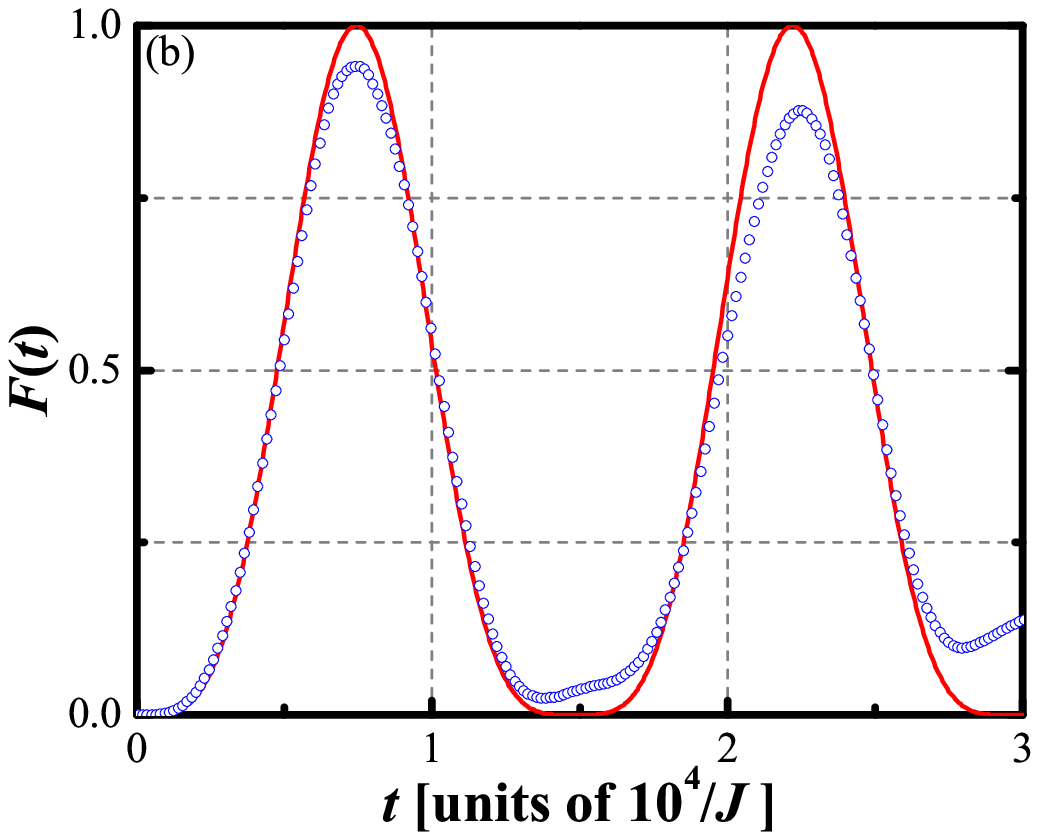}
\caption{(Color online) Comparison of the exact transition probability $F(t)$
(open circle) with the analytic result of Eq. (27) (red solid line) for the
system with $N=499$, $l=2$, $J_{0}=2\times 10^{-3}J$, $\protect\mu _{0}=0.1J$
(upper plot), and $0.05J$ (lower plot). Time is expressed in units of $%
10^{4}/J$. It shows that, small $J_{0}$ leads to the result about transfer
fidelity is very close to that described by the effective Hamiltonian $H_{%
\text{eff}}$. }
\label{fig4}
\end{figure}

\subsection{Robustness to Disorder}

We now turn to the performance of spin chains in the presence of static
imperfections in the couplings, which are unavoidable in experimental
implementations. We will show that the energy gap can protect the
performance of the QST in the presence of static disorder in the system
parameters.

We now assume that the tunnel coupling of the medium Hamiltonian has a
random but constant offset, $\delta \epsilon _{j}$, i.e.,%
\begin{eqnarray}
\hat{H}^{\prime } &=&\sum_{j=1}^{N-1}-J(1+\delta \epsilon _{j})\left( \hat{a}%
_{j}^{\dag }\hat{a}_{j+1}+\text{h.c.}\right)  \notag \\
&&-\mu _{0}\hat{a}_{N_{0}}^{\dag }\hat{a}_{N_{0}}-\mu \left( \hat{a}%
_{L}^{\dag }\hat{a}_{L}+\hat{a}_{R}^{\dag }\hat{a}_{R}\right) -J_{0}\left(
\hat{a}_{L}^{\dagger }\hat{a}_{N_{0}-l}+\hat{a}_{R}^{\dagger }\hat{a}%
_{N_{0}+l}+\text{h.c.}\right) ,
\end{eqnarray}%
where $\delta $ is the maximum coupling offset bias relative to $J$; $%
\epsilon _{j}$ is drawn from the standard uniform distribution in the
interval $\left[ -1,1\right] $ and all $\epsilon _{j}$ are completely
uncorrelated with all sites along the chain.

\begin{figure*}[tbp]
\center
\includegraphics[bb=83 469 383 706, width=6.3 cm, clip]{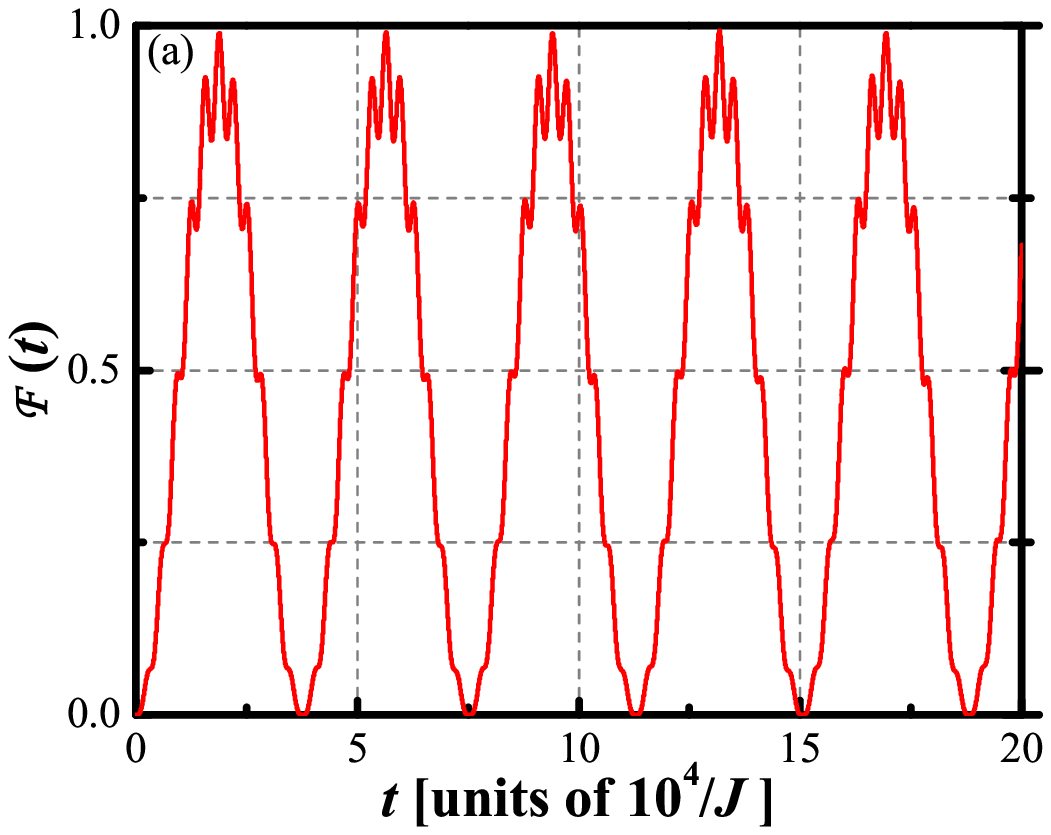} %
\includegraphics[bb=114 469 383 706, width=5.65 cm, clip]{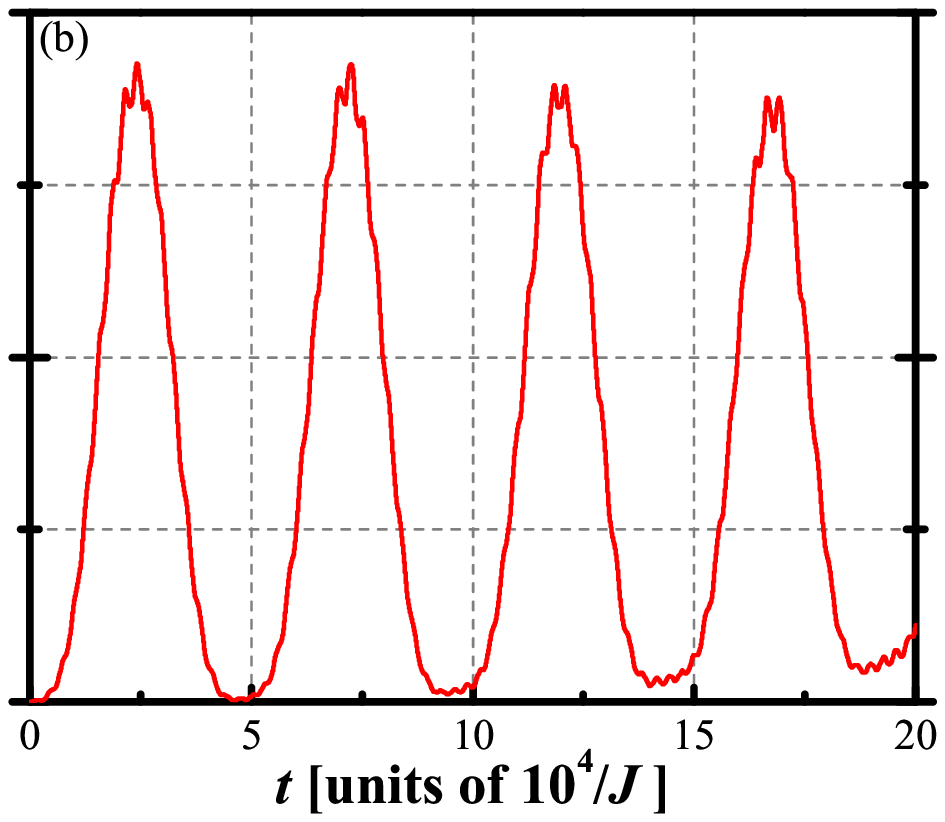} %
\includegraphics[bb=114 469 383 706, width=5.65 cm, clip]{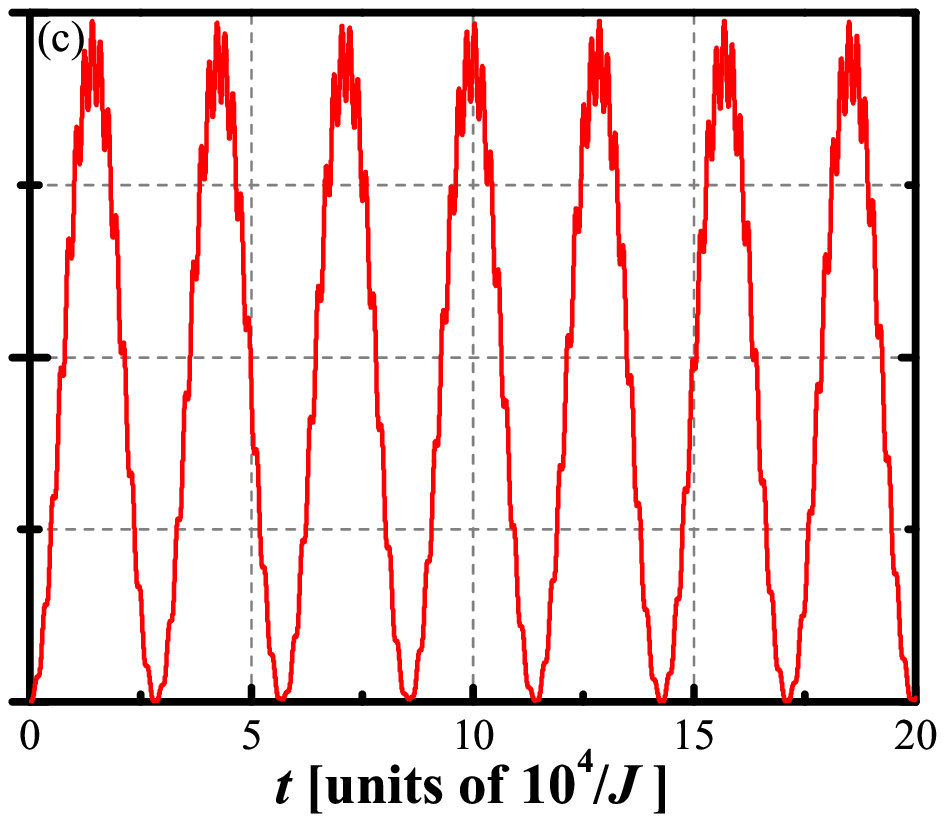}
\caption{(Color online) The transition probability $F(t)$ of QST as a
function of time in a $N=499$ system with $J_{0}=2\times 10^{-3}J$, $l=2$
and random imperfections of coupling strength $\protect\delta $ function of
time. The left figure corresponds to the case when $\protect\mu _{0}=0.1J$
and $\protect\delta =5\times 10^{-3}$, the middle figure to $\protect\mu %
_{0}=0.1J$ and $\protect\delta =1\times 10^{-2}$, and the right figure to $%
\protect\mu _{0}=0.5J$ and $\protect\delta =1\times 10^{-2}$. The results
shows that weak fluctuations in the coupling strengths do not deteriorate
the quality of QST due to the exitance of energy gap.}
\label{fig5}
\end{figure*}

We numerically calculate the Schr\"{o}dinger equation for the dynamical
evolution and compute the overlap, $\mathcal{F}(t)=\left\vert \left\langle
R\right\vert e^{-i\hat{H}^{\prime }t}\left\vert L\right\rangle \right\vert
^{2}$, to assess the performance of the chain. In Fig.~\ref{fig5} we plot
the behavior of $\mathcal{F}(t)$\ as a function of time, $t$, in the system
with $N=499$ QDs, $l=2$, $J_{0}=2\times 10^{-3}J$ for three cases: (a) $\mu
_{0}=0.1J$, $\delta =5\times 10^{-3}$, (b) $\mu _{0}=0.1J$, $\delta =1\times
10^{-2}$, and (c) $\mu _{0}=0.5J$, $\delta =1\times 10^{-2}$. From this
comparison, one can see that (i) this scheme is robust against the static
disorders that would be unavoidable in experimental implementations and (ii)
that the large energy gap (or large $\mu _{0}$) is more robust than small
one against disorder.

\section{SUMMARY}

According to quantum mechanics, it is not difficult to establish a
long-distance QST using a gap system. However, the magnitude of the gap in
this kind of scheme is crucial: first, the gap should be independent of the
size of the system; second, the energy gap should be manipulated as required
for perfect QST. The reason is that if the gap is too large, the QST period
increases exponentially with the distance between two distant parities; when
the gap is too small, the fidelity of the QST is reduced.

In this study, the quantum transmission of an electron through an IGS
(serving as the data bus) is studied by theoretical analysis and numerical
simulation. First, we show that the IGS has a nonvanishing energy gap above
the ground state, which depends only on the on-site energy, $-\mu _{0}$, of
the impurity. The approach to realize perfect QST is based on weakly
connecting two external QDs with the bus. Different transfer distances can
be achieved by suitable choices of connecting sites to the data bus. By
treating the weak coupling as a perturbation, we find that a gap system can
induce an effective three-level Hamiltonian [Eq.~(\ref{H3})]. This
theoretical result is confirmed by performing numerical simulations;
moreover, the effective coupling, $J_{\text{eff}}$, also decays slowly with
increasing transfer distance if the system parameters are chosen reasonably.

Furthermore, the fault tolerance for more realistic system parameters is
also demonstrated. It has been shown that perfect state transfer can also be
achieved in the presence of disorder. For larger values of the energy gap
(or $\mu _{0}$), the effect of disorder on the quality of QST will be
strongly suppressed.

\section*{ACKNOWLEDGEMENTS}

Z. Song thanks the support of the National Basic Research Program (973
Program) of China under Grant No. 2012CB921900. We acknowledge the supports
of the National Natural Science Foundation of China (Grant Nos. 11105086,
11174027, 11374163, 11121403, 10935010, and 11074261).

\end{document}